\documentclass[preprint,aps]{revtex4}
\usepackage{graphics}
\usepackage{amssymb}

\def\qed{{\hfill $\Box$}}

\begin{document}
\title{Low-Pass Filters: Commentary on a Remark by
Feynman}
\author{A.G. Ramm}
\affiliation{Mathematics Department\\
Kansas State University\\Manhattan, KS 66506} \email{
ramm@math.ksu.edu}
\author{L. Weaver}
\affiliation{Physics Department\\Kansas State University\\
Manhattan, KS 66506} \email{lweaver@phys.ksu.edu}
%\date{}

\begin{abstract}
In Feynman's lectures there is a remark about the limiting value
of the impedance of an n-section ladder consisting of purely
reactive elements (capacitances and inductances). The remark is
that this limiting impedance $z=\lim_{n\rightarrow\infty}z_n$ has
a positive real part. He notes that this is surprising since the
real part of each $z_n$ is zero, therefore it is impossible for
the limit to have a positive real part. A recent article in this
journal offered an explanation of this paradox based on the fact
that realistic impedances have a non-negative real part, but the
authors noted that their argument was incomplete. We use the same
physical idea, but give a simple argument which shows that the
sequence $z_n$ converges like a geometric series. We also
calculate the finite speed at which energy is propagated out into
the infinite ladder.
\end{abstract}
\maketitle

\section{Introduction}

In Feynman's lectures \cite{fey64} there is a remark about the
limiting value of the impedance of an n-section ladder consisting
of purely reactive elements (capacitances and inductances). He
shows that the impedance obeys the recurrence relation
\begin{equation}
z_{n+1}=Z_1 + 1/(1/Z_2  + 1/z_n)
\end{equation}
If $\lim_{n\to \infty}z_n := z$ exists then $z$ is the solution of
\begin{equation} z=Z_1 + 1/(1/Z_2  + 1/z)
\end{equation} and this limiting impedance has a positive real part. Feynman
points out that this is surprising since the real part of each
$z_n$ is zero, therefore it is impossible for the limit to have a
positive real part. A recent article \cite{kri03} in this journal
discussed this paradox as an example of the need for care when
taking limits.

A comment \cite{ure03} on [2] tells us that there is no need to
worry about taking limits because there are ways to define
infinite series other than as the limit of a sequence of finite
series. The author asserts that we should just define the
impedance of the infinite ladder to be one of the solutions to
equation (2) above. This may be considered as a {\it definition}
of the impedance of the infinite ladder, but does nothing to
remove the surprise we feel that the properties of the so-defined
infinite ladder are so different from the properties of $z_n$ for
large $n$. The feeling of surprise is even more acute in the
example given in [3]. This example is the geometric series
$Z_n=\sum_{j=0}^np^jR$ which arises in an infinite ladder of
increasing resistances. When $n$ tends to infinity and $|p|<1$ the
sum of this series is $Z=R/(1-p)$, and in [3] we are told to use
this form even if $p>1$ on the grounds that this is the Borel sum
of the series. But this does not remove our astonishment that,
according to [3], $1+2+4+8+...=-1$, taking $p=2$ for example.
Furthermore, contrary to [3], when $Re\, p >1$ the Borel sum of $
\sum_{j=0}^{\infty} p^jR $ does not exist \cite{ww1927}.

The comment [3] gives another argument that is closer to
Feynman's. This is that the impedance $Z(R)$ of the infinite set
or resistances should obey $Z-R=pZ$, on the grounds that removing
the first element of the series leaves still an infinite series.
But if we remove the first element of the ladder we get an
infinite ladder whose first element is $pR$, not $R$. Thus the
recursion relation should be $Z(R)-R=Z(pR)$, and the only
rationale we see for $Z(pR)=pZ(R)$ is that what we mean by $Z(R)$
is $Z(R)=\lim_{n\to\infty}Z_n(R)$. Thus again we need to consider
the usual limit, which [3] rejects. Finally the author of [3]
points out that we can consider the infinite series as a finite
series, appropriately terminated. If we terminate the $n$-section
ladder with an active element having resistance $-p^{n+1}R/(p-1)$
then we get $Z=R/(1-p)$ as he wants. But why this value? Why not
$-p^{n+1}R/(p-1)+\lambda R$? Then the infinite sum is different.
(Another possible rationale, though not one offered in [3], is
this: The $Z_n$ obey the recursion relation $Z_{n+1}=R+pZ_n$. This
gives $Z_{n+1}=R/(1-p)+p^n(Z_1-R/(1-p))$. This converges to
$R/(1-p)$ for $p<1$ and for $p>1$ it converges to the same value
{\it if} $Z_1=R/(1-p)$. But then {\it every} $Z_n$ equals $Z_1$.
This does not seem to be the physical situation envisioned in
[3].)

For these reasons we disagree with [3] and think the authors of
[2] are right to wonder how to resolve the puzzle noted by Feynman
in [1].

So we return to the recursion relation, equation (1), and
investigate its convergence. We use the notation of [2]. The
authors of [2] suggested that we consider each $Z_1, Z_2$ to have
a small imaginary part $r$ and defined the impedance of the
infinite ladder to be the iterated limit $\lim_{r\to 0} \lim_{n\to
\infty}z_n$, noting that the order in which the limits are taken
is crucial. They then gave an argument that this limit exists,
based on the contraction mapping principle. Unfortunately their
proof is incomplete, as they note, and the reader may be misled
about the proper use of the principle. We give here a rigorous
proof based on elementary reasoning suitable for an undergraduate
classroom and then make some remarks on the contraction mapping
principle. We also show that energy is propagated out into the
infinite network with a finite velocity which we calculate.

\section{EXISTENCE OF THE LIMITING IMPEDANCE}

The notation is simplified by defining

\begin{equation}
p_n=z_n/Z_1, \quad p_{\pm}=z_{\pm}/Z_1,\quad\hbox{and}\quad
t=Z_2/Z_1 ,
\end{equation}
where $z_{\pm}$ are the solutions of equation (2). \noindent  Then
 equation (1) becomes
 \begin{equation}p_{n+1} = 1+  tp_n/(t+p_n),\end{equation}
 the limiting value $p=p_+$ or $p_-$ obeys  $p^2=p+t$, and
$p_{\pm}=(1/2)(1 \pm \sqrt{1 + 4t})$. Define the branch of the
square root by $\sqrt{1 +4t} = a+ib$ with $a>0$.  This is always
possible when $t$ is not on the negative real axis and less than
$-1/4$. ( $t\not{\!\!\in} (-\infty,-1/4)$.) That is, we have cut
the complex $t$-plane from $-1/4$ to $-\infty$ and all our work is
done on one sheet of the Riemann surface.
 Our first goal is to show that $\lim_{n\rightarrow\infty}p_n=p_+$.
 We will see that $p_-$ is an unstable fixed point. So consider now
 $p_1\ne p_-$.

{\bf Theorem:} {\it For $p_1 \ne p_-$ and $Re \sqrt{1+4t} >0$ the
sequence  defined by}

\begin{equation}
p_{n+1} = 1+ tp_n/(t+p_n)
\end{equation}

\noindent {\it converges to $p_+$ at the rate of a geometric
series.}

{\bf Proof:} It is easily checked that

\begin{equation}
{p_{n+1}-p_+\over p_{n+1}-p_-} = ({p_nt\over p_n+t} -  {p_+t\over
p_++t})/({p_nt\over p_n+t} - {p_-t\over p_-+t})={p_n-p_+\over
p_n-p_-}\cdot{p_-+t\over  p_++t}={p_n-p_+\over p_n-p_-}\cdot
{p_-^2\over p_+^2}.
\end{equation}

\noindent Let $c_n:={p_n-p_+\over p_n-p_-}$ and $\gamma
:={p_-\over p_+}$.   Then equation (6) implies

\begin{equation}
|c_{n+1}|=|c_n||\gamma|^2,
\end{equation}

\noindent and $|\gamma|^2 = ((a-1)^2+b^2)/((a+1)^2+b^2)$ is less
than $1$ {\it if and only if}  $a>0$. Thus
$|c_n|=|\gamma|^{2n}|c_1|\rightarrow 0$ as $n\to\infty$ and so
$p_n\to p_+$. This shows that the limit $\lim_{n\to\infty}p_n$
{\it does exist} for all $p_1 \ne p_-$, it has positive real part,
and the convergence is that of a geometric series. \qed

{\it Remark 1:} {\it If $p_1=p_-$, then $p_n=p_-$ for all $n$, so
the limit $\lim_{n\to\infty}p_n=p_-$ does exist in this case also.
However, an arbitrarily small perturbation of the initial data
$p_1=p_-$ results in the limit $\lim_{n\to\infty}p_n=p_+$. This
means that $p_+$ is the stable fixed point for the iterative
process (5), while $p_-$ is an unstable fixed point for this
process.}

{\it Remark 2:} {\it There is a geometrical interpretation of the
above proof which explains the reasons for formula (6) to hold.
Namely, the mapping $p \to c(p):= (p-p_+)/(p-p_-)$ maps the
complex plane onto itself, taking $p_+$ to the origin and $p_-$ to
the point at infinity. In the complex  $c$-plane the origin is a
global attractor.}

Apply this now to the low-pass filter discussed by Feynman. In
this case $Z_1=i\omega L+r$, $Z_2=1/(i\omega C) +r'$ with $r,r'$
small and positive, so

\begin{equation}
\begin{array}{lllll}
1+4t &= &+A-i\epsilon, A>0, \; &\hbox{\rm for} \; &\omega >\omega_c\\
     &= &-A-i\epsilon, A>0, \; &\hbox{\rm for} \; &\omega < \omega_c.
\end{array}
\end{equation}

\noindent In these formulas $\epsilon=O(r+r^\prime)$ as
$r,r^\prime\rightarrow0$, so $\epsilon$ is a small positive
number,
 and $A > 0$ is $A
=|1-{\omega_c^2/\omega ^2}|$. The cut-off frequency
$\omega_c=2/\sqrt{LC}$. Choosing the branch of the square root
with positive real part gives

\begin{equation}
\begin{array}{lllll}
\sqrt{1+4t} &= &+\sqrt{A}-(i\epsilon /2)\sqrt{A}+ O(\epsilon
^2)\; &\hbox{\rm for}\; &\omega >\omega_c \\
{\;}&= &-i\sqrt A+(\epsilon /2)\sqrt{A}+ O(\epsilon ^2)
\;&\hbox{\rm for}\; &\omega < \omega_c.
\end{array}
\end{equation}

\noindent Here and below $\sqrt{A}>0$.  Rewrite this now in terms
of the limiting impedance $z_+=Z_1p_+ = (Z_1/2)(1 + \sqrt{1
+4t})$. For $\omega < \omega_c$ we find

\begin{equation}
z_+=[(i\omega L +r)/2][1-i\sqrt A + (\epsilon /2)\sqrt A +
O(\epsilon^2)] =  (\omega L/2)\sqrt A +i\omega L/2 + O(\epsilon).
\end{equation}

\noindent The  real part of $z_+$ is positive, consistent with the
principle that the real part of the impedance of a passive network
cannot be negative. We can now take the limit $\epsilon
\rightarrow 0$ and get $\lim_{\epsilon\rightarrow 0}z_+=(\omega
L/2) \; (\sqrt{A}+i)$.
 For $\omega > \omega_c$ we find

\begin{equation}
z_+= [(i\omega  L + r )/2][1+\sqrt A-(i\epsilon/2)\sqrt A
+O(\epsilon^2)]=(i\omega  L/2)(1+\sqrt A) +
 O(\epsilon).
\end{equation}

\noindent Again, the real part is positive (so it  is physically
reasonable), and we can take the limit $r,r'\rightarrow 0$ and
find $\lim_{\epsilon\rightarrow 0}z_+= (i\omega L/2)(\sqrt{A}+1)$.
Taking the limit $\epsilon\rightarrow 0$ is equivalent to taking
$r,r^\prime \rightarrow 0$.

We have shown that $z_+=\lim_{n\to\infty}z_n$ {\it does exist}
when the impedances have a positive real part. {\it The
requirement needed to make the mathematics rigorous is exactly the
requirement that makes physical sense: the real part of a
physically realistic passive impedance must be positive.} Once
that is recognized one can describe the physical idealization of
perfect impedances by taking the limit as these real parts go to
zero. This gives exactly the result obtained by Feynman. We have
treated a low-pass filter here, but the same analysis applies to a
high-pass filter in which the inductance and capacitance are
interchanged.

Now that we know that the limits make sense we can discuss the
infinite network further. If the infinite ladder is connected to a
source, and the source is modulated in time, then the modulation
propagates out into the ladder like a wave with well-defined group
velocity. Thus energy can be delivered by the source out into the
infinite network and the energy does not come back, it is
absorbed: the infinite network acts like a "black box" whose
impedance has a positive real part.

To see this, first use Kirchhoff's laws to show that
$\tilde{V_n}(\omega)=(-\gamma)^n {\tilde V}_s(\omega)$ as done in
[1]. Then put $-\gamma = e^{i\delta(\omega)}$. Now describe the
source voltage $V_s(t)$ by the modulated signal
$V_s(t)=f(t)e^{i\omega_0 t}$, where the modulating function $f$ is
slowly varying. The Fourier transform of the source voltage is
$\tilde{V_s}(\omega)= \tilde {f}(\nu)$, where
$\nu:=\omega-\omega_0$, and the spectrum of $f$ vanishes outside a
small neighborhood $|\nu|< \nu_0$ of the zero frequency. Then the
time-dependence of the voltage on the $n-$th section of the ladder
is $V_n(t)=e^{i\omega_0 t}{1\over {2\pi}}\int_{|\nu|<\nu_0}d\nu
e^{i\nu t +in \delta(\omega_0+\nu)}\tilde {f}(\nu)$.  Compare this
with the standard representation of a wave packet
$g(x,t):={1\over{2\pi}}\int_{|\nu|<\nu_0}d\nu e^{i\nu
t-ixk(\nu)}\tilde{f}(\nu)$, with group velocity $v:={d\nu\over
{dk}}|_{\nu=0}$. In our case the role of the variable $x$ is
played by parameter $n$ and $k=-\delta(\omega_0+\nu)$, so
$v=-1/\tau$, where $\tau:={d\delta(\omega)\over
d\omega}|_{\omega=\omega_0}>0$. The quantity $\tau$ has the
physical meaning of the time needed for the wave to propagate
through one section of the ladder.

\section{THE CONTRACTION MAPPING PRINCIPLE}

In [2] the authors propose to apply the contraction mapping
principle to prove the convergence of $z_n$ for $\omega <
\omega_c$, where

\begin{equation}
z_{n+1}=f(z_n), \quad z_1=Z_1 +Z_2,
\end{equation}

and

\begin{equation}
f(z)=Z_1 + 1/(1/Z_2 + 1/z).
\end{equation}

\noindent They argue that the sequence converges to a fixed point
$\zeta =f(\zeta)$ only if the mapping in equation (7) is a
contraction mapping ([2], line 1 below (5)). This is not right.
Let us review the contraction mapping principle \cite{bag1992}.

The contraction mapping principle says:
 \noindent
{\it If there is a closed set $D$ such that (a) $f(D)\subseteq D$
and (b) $|f(z')-f(z)|\leq q|z'-z|$  for all $z,z'\in D$ with $q$ a
constant, $0<q<1$, then there is in $D$ a unique fixed point
$\zeta=f(\zeta )$ of the map $f: D\rightarrow D$ and the sequence
$z_{n+1}=f(z_n), z_1 \in D$, converges to $\zeta$}.

In [2] the authors do not specify a $D$ that is mapped into itself
by $f$ nor do they check that their initial approximation $z_1$
generates a sequence that ever reaches $D$. They recognize this
gap and point it out in a footnote. They have simply calculated
$|f'(\zeta)|$ for the fixed points $\zeta$ and say that only if
$|f'(\zeta)|<1$ will the sequence converge. This is false in
general: sequences not generated by contraction maps may converge.
A simple example is $f(z)=z^2 +1/4$. The sequence $z_{n+1}=f(z_n)$
converges to the fixed point $\zeta=1/2$ for $z_1$ in $D=[0,1/2]$
but $f'(1/2)=1$, and although $f(D)\subseteq D$, this $f$ is not a
contraction mapping because there is no $q<1$ that satisfies
condition (b) above for this set $D$. Furthermore, fixed points
are possible even when $f'>1$. A nice example of this is $f(z) =
\tan (z)$: at all the fixed points $\zeta_j=\tan \zeta_j$ we have
$|f'(\zeta_j)|=|\sec ^2\zeta_j| \ge 1$.

In section 2 we proved convergence of the sequence in equation (1)
without appeal to the contraction mapping principle. Our proof
shows that if the conditions specified there are satisfied, then
there exists a closed set $D_\epsilon=\{z:|z-z_+|\leq\epsilon\}$
such that for sufficiently small $\epsilon >0$ the map $f$ does
map $D_\epsilon$ into itself and is a contraction. The initial
approximation $z_1=Z_1+Z_2$ does not belong to this set, but our
proof shows that for any $z_1 \ne z_-$ the approximations do
eventually reach $D_\epsilon$, as the authors of [2] hoped.

\end{document}